\documentclass[12pt]{article}
\usepackage{epsf}
\begin{document}
\begin{titlepage}
\begin{flushright}
TPI-MINN-99/41-T  \\
UMN-TH-1816/99   \\
ITEP-TH-44/99\\
hep-th/9910071 \\
\end{flushright}

\vspace{0.6cm}

\begin{center}
\Large{{\bf BPS Saturated Solitons in ${\cal N}=2$ Two-Dimensional 
Theories
on
$R\times S$ (Domain Walls in Theories with  Compactified
Dimensions)}}

\vspace{1cm}

Xinrui Hou$^{1)}$,  A. Losev$^{1,2)}$, and M. Shifman$^{1)}$
 \end{center}
\vspace{0.3cm}

\begin{center}
$^1$  {\em Theoretical Physics Institute, Univ. of Minnesota, 
Minneapolis,
       MN 55455}

$^2$ {\em Institute of Theoretical and Experimental Physics, Moscow 
117259,
Russia}$^\dagger$
 \end{center}

\vspace{0.5cm}

\begin{abstract}
We discuss topologically stable solitons in two-dimensional theories 
with the
extended supersymmetry assuming that the spatial coordinate is 
compact. 
This
problem arises in the consideration of the domain walls in the 
popular 
theories
with compactified extra dimensions. Contrary to  naive expectations, 
it 
is shown
that the solitons on the cylinder can be BPS saturated. In the case of 
one
chiral superfield, a complete theory of the BPS saturated solitons is 
worked
out. We describe the classical solutions of the BPS equations. 
Depending 
on the
choice of the K\"ahler metric, the number of such solutions can be 
arbitrarily
large. Although the property of the BPS saturation is preserved order 
by 
order
in perturbation theory, nonperturbative effects eliminate the 
majority 
of the
classical BPS states upon passing to the quantum level. The number 
of 
the
quantum BPS states is found. It is shown that the ${\cal N}=2$ field 
theory
includes an auxiliary ${\cal N}=1$ quantum mechanics, Witten's 
index
of which counts the number of the BPS particles.
\end{abstract}

\vspace{0.5cm}

\begin{flushleft}
$^\dagger$ Permanent address
\end{flushleft}
\end{titlepage}

\section{Introduction and Physical Motivation}
\label{sec1}

The idea that our matter is made of  zero modes trapped on the
surface of a (1+3)-dimensional topological defect (domain wall)
embedded in a higher-dimension universe dates back to
1983~\cite{a1}. With the advent of
supersymmetry and Planckian physics, it was natural to attempt to
exploit~\cite{a2} the setup for  solving  a wide range of questions,
such as building a fundamental theory at the scale much below 
Planckian,
developing an appropriate pattern of supersymmetry breaking,
and so on. The BPS saturated
domain walls which preserve a part of the original supersymmetry, 
play
a special role. It was shown~\cite{a2,a3} that the scalar and spinor
matter, as well as the
gauge fields, can be localized on (1+3)-dimensional dynamical walls.
The next step was made in~\cite{a4} where gravity was included in
consideration.  Since the gravitons are not localizable on dynamical
walls, it
was suggested that the original multidimensional space-time is 
compact
with
respect to one or more coordinates, it has the structure of a cylinder
$S^{k}\times {\cal M}_{4}$ ($k\geq 1$), and the gravitons propagate 
in 
the
bulk. All
other fields are still localized on the wall that appears dynamically 
on
the
cylinder. The approach was later dubbed {\em large extra 
dimension(s)
theories}. Phenomenologically it is preferrable~\cite{a4} to have 
$k=2$,
the wall
width of order, roughly, 1 (TeV)$^{-1}$, and the radius of the
cylinder
of order
1 mm. For a recent discussion of the emerging quite rich 
phenomenology
see
e.g.~\cite{a5}.

Surprisingly, dynamical aspects of this construction have been
 investigated at a
rather fragmentary level. In particular, the question of interest is the
issue
of the BPS saturation of the wall-like topological defects on the 
cylinder,
which has never been addressed in full previously. Studying such
topological
objects is the task of this work.
Since the domain wall is a static
field configuration depending on a single coordinate,
while other spatial coordinates are passive, this problem
in many aspects is equivalent to studying BPS saturated solitons
in two-dimensional theories obtained
by dimensional reduction of multidimensional theories. In other 
words,
one starts from $D$-dimensional theory, in which $D-1$ coordinates
are spatial. The wall solution depends on one of them.
 The presence of ``extra"
$D-2$ spatial dimensions is  irrelevant at the classical level (although 
it
may
be relevant at  the quantum level).  The extra $D-2$  dimensions
can be reduced. The two-dimensional theory obtained in this way
has an extended supersymmetry (SUSY).  We will consider here the
generalized Wess-Zumino models, also referred to as the
Landau-Ginzburg theories, in two dimensions.

For instance, if one starts from   (1+3)-dimensional
Wess-Zumino model with a discrete set of SUSY vacua,  the walls
are two-dimensional objects (two space dimensions plus time) which 
can
be assumed to lie in the
$yz$ plane. The wall profile depends  on $x$.
If  the original theory is considered in a noncompact space,
${\cal M}_{4}$, the  topological
stability of the wall  is achieved in a rather trivial manner.
Let $\Phi_{*1}$ and
$\Phi_{*2}$ be two distinct degenerate vacua
of the theory, $V(\Phi_{*1})= V(\Phi_{*2})=0$, where $V$ is the  
scalar
potential. Then, the minimal
energy solution $\Phi(x)$ interpolating between $\Phi_{*1}$ at $x=-
\infty$ and $\Phi_{*2}$ at $x=+\infty$ is  topologically stable.
(Here $x$ is the spatial coordinate.) Such
domain walls (solitons) always exist. They may or may not be BPS
saturated.  The issue of the BPS saturation requires a separate
dynamical
consideration~\cite{a6}. A general theory of the BPS saturated 
solitons
in the ${\cal N}=2$ Landau-Ginzburg models in this case was worked 
out
in~\cite{Cecotti:1993rm}.

If the world sheet is a  cylinder, $R\times S$, the situation with the
topological
stability is different. Indeed,
$x$ now becomes a compact coordinate. If the radius of $S$ is $r$,
the points $x$ and $x+2\pi r$ are identified (we will also use the
notation
$L=2\pi r$ and will often put $r= (2\pi)^{-1}$ in what follows,
so that $L=1$). To produce
topologically
stable solitons, the dynamical theory under consideration
must be such that the field
$\Phi$ is defined  on a manifold $ M$ with a noncontractible cycle
or
several distinct cycles. If, as one winds around $S$, i.e. $x$ varies
continuously
from $x=0$ to $x=L$, the field $\Phi$ winds around a cycle of $M$,
the corresponding field configuration $\Phi (x)$ will be topologically
stable~\cite{a2}. Unlike the theories on noncompact manifolds,
 the topologically
stable configuration described above need not be related to the 
vacua of
the theory. The latter may not exist at all (the run-away theories). 
Thus,
the  issue reduces to the classification of noncontractible closed 
contours 
\footnote{The space of
all loops in $M$ could be divided into classes such
that loops inside each class can be smoothly deformed
into each other. Such space of classes is denoted by
$\pi_1(M)$.}
 on $ M$, i.e. the fundamental group $\pi_1( M)$.

Let us assume that a topologically stable soliton exists.  Can it be BPS
saturated?
As is well-known, the issue of the saturation is related to the 
existence
of the central
charge(s) $Z$ in the superalgebra~\cite{Witten:1978mh}. The
nonvanishing central charge is a necessary condition for the soliton 
to 
be
saturated.
Since the spatial coordinate is compact in our problem,
$\Phi (0) = \Phi (L)$, and the superpotential $W$ is a holomorphic
function of $\Phi$,  at first sight one might conclude that
$Z = \left| W[\Phi (0)] - W[\Phi (L)]\right| =0$,  and the saturated
solitons are
not  possible. In fact, this conclusion is wrong. It is the differential
$dW$ that must be  a
single-valued function (it determines the scalar potential),
the superpotential need not be single-valued.
The superpotential $W$ can be  a locally holomorphic function with 
branches.
Then the central charges are
determined by the integrals of $dW$ over various noncontractible 
cycles 
on
$ M$, i.e. by periods of the differential $dW$,
\begin{equation}
Z_i = \int_{{\rm nc\,\, cycle}_i}\, d W\, .
\label{incc}
\end{equation}

The nonvanishing central charge is a necessary but not sufficient
condition
for the existence of the BPS saturated soliton.
We need to make two more steps.
First, we need to look for the static classical configurations that
satisfy the BPS equations.
Then, we need to take into account quantum effects.
If the classical BPS configuration is isolated,
one can construct a quasiclassical state
around such configuration.
The perturbative quantum corrections would change the shape of 
this
quasiclassical state, but
at the level of perturbative quantization, the
modified states would still be annihilated by half of the 
supercharges.
At the same time, due to nonperturbative quantum effects, not all 
these
states  may survive in the nonperturbative quantum theory  --
due to instanton-like phenomena, the half of the supercharges that 
acted
trivially at the classical level, may in fact start acting nontrivially,
connecting these states.
Then, such states cease to be BPS saturated. This phenomenon was 
first 
observed
by Witten~\cite{Witten:1982dg} in the interpretation
of the Morse theory in terms of ${\cal N}=1$ supersymmetric 
quantum 
mechanics.
The critical points of the superpotential were quasiclassical
``BPS configurations",
and instantons  provide mixings between them.

The paper is organized as follows.
In Sec. 2 we will study
the space of the classical BPS configurations
and present a heuristic
 evaluation of the number of
the BPS states in the generalized Landau-Ginzburg models
based on the properties of
Cecotti--Fendley--Intriligator--Vafa (CFIV) 
index~\cite{Cecotti:1992qh}.
We will show that only vicinities of the poles
of $dW$ contribute to the BPS solitons at the quantum level.

In the Sec. 3 we will illustrate the general consideration
of Sec. 2 by a typical example.
In the Sec. 4 we clarify the assertions made in Sec.
2, without the use of the CFIV index. We show that the
 ${\cal N}=2$ supersymmetric
field theory contains within it an auxiliary ${\cal N}=1$
quantum mechanics, so that  Witten's index of the latter
counts the number of the BPS solitonic states in the former.
{\it En route} we show that the
``localization to the poles of $dW$"  phenomenon has analogs
in the ${\cal N}=2$ supersymmetric quantum mechanics
on the space with freely acting isometries (the nonvanishing Killing 
vectors).

\section{A Heuristic Derivation
of the Number of BPS States \,in \,the \,Generalized \,
Landau--Ginzburg\, 
\\Model}

\subsection{Generalized Landau-Ginzburg Model and the
Classical BPS Equations}

The action of the generalized Landau-Ginzburg (GLG) model,
to be considered below, has the form
\begin{equation}
S = \frac{1}{4}\int\! {\rm d}^2 x \,{\rm d}^4\theta \,
 { K}(\Phi^i ,
\bar{\Phi}^j )+
\left\{ \frac{1}{2}\int\! {\rm d}^2 x \,{\rm d}^2\theta \,
{ W}(\Phi^i)
+{\rm
H.c.}\right\}\, ,
\label{klagrwz}
\end{equation}
where $x^\mu = \{ it, x\}\quad (\mu = 0,1)$ and $x\in S$.
Moreover, $\Phi^i$ is a set of the chiral superfields
(corresponding to the holomorphic coordinates on the target space
$M$), the
superpotential
${ W}$ is a (multivalued) analytic function of all chiral
 variables $\Phi^i$,
while the kinetic term is determined by the  K\"{a}hler
 potential   ${ K}$,
which is  a real (multivalued) function
depending
both on chiral $\Phi^i$ and antichiral $\bar \Phi^{\bar{j}}$ fields.

While $W$ and $K$ are multivalued, the K\"ahler metric,
 \begin{equation}
G_{i\bar j}=\frac{\partial^2\,{ K}}{\partial \Phi^i
\partial \bar \Phi^{\bar
j}}
\label{metric}
\end{equation}
and the 1-differential  $\Omega=\Omega_i d\Phi^i$,
\begin{equation}
\Omega_i(\Phi)= \frac{\partial W}{\partial \Phi^i}
\end{equation}
are single-valued on the target space $M$. This is a necessary 
requirement, to
be imposed here and below. It ensures that the scalar potential and 
the 
fermion
terms in the Lagrangian are well-defined.

In components the Lagrangian takes the form
\begin{equation}
{\cal L} =\sum_{i,\bar{j}=1}^n \left\{G_{i\bar j}\,
\partial_\mu\Phi^i\,\partial_\mu\bar\Phi^{\bar j} +  G^{i\bar j}
\frac{\partial { W}}{\partial \Phi^i }\,\frac{\partial
 \bar{W}}{\partial
\bar\Phi^{\bar j} } \right\} + \mbox{fermions}\,,
\end{equation}
where $n$ is the number of the chiral (antichiral) fields involved, 
and
 $G^{i\bar j}$ is the inverse matrix,
$$
G_{i\bar j} G^{k\bar j}= \delta^k_i\, .
$$

 The equations of the
BPS saturation have the form
\begin{equation}
\dot{\Phi}^i = e^{i\delta}\,  G^{i\bar j}\,  \frac{\partial\bar
W}{\partial\bar\Phi^{\bar j}}\, ,
\quad \dot{\bar{\Phi}}^{\bar j} = e^{-i\delta} \,
G^{i\bar j}\, \frac{\partial W}{\partial\Phi^i}\,,
\label{bpseq}
\end{equation}
where the dot  denotes differentiation over the {\em spatial}
coordinate $x$.
Let us denote by $\Gamma \in M$ the loop in $M$ that is the
image of the map $\Phi$.
Then,  the phase $\delta$ appearing in Eq. (\ref{bpseq}) is that of the
period,\footnote{The period integral depends only on the class of
$\Gamma$ in $\pi_1(M)$.}
\begin{equation}
\int_{\Gamma}\, \Omega \equiv \Delta_i W \equiv e^{i\delta }\left|
\Delta_i W \right|\, .
\label{oprf}
\end{equation}
The formula (\ref{bpseq}) must be viewed as the master equation.

\subsection{How to Solve the BPS Equations for the 
One-Dimensi\-onal Target Space}

The general solution of Eq. (\ref{bpseq}) for the one-dimensional
target space
could be divided into two steps.
At the first step we will find the space
${\cal S}$ of solutions $\Phi(x,s)$
of Eq. (\ref{bpseq}) ($s$ is the coordinate on the space ${\cal S}$),
which satisfy a modified periodicity condition,
\begin{equation} \label{ls}
\Phi(0,s)=\Phi(l(s),s)\,,
\end{equation}
for some function $l(s)$.

Then, we will pick up the proper elements from ${\cal S}$, the
classical BPS solitons,
by imposing the condition
\begin{equation} \label{lsl}
l(s)=L\,.
\end{equation}
The requirement (\ref{lsl}) determines (generally speaking) a 
discrete
set of the parameters $\{s_i\}$
for which one deals with the classical BPS solutions.

Let us start from the first step.
The space ${\cal S}$ could be considered as
a space of such closed unparametrized curves on $M$
that:

(i)  These curves are
tangent to the vector field with the components
(Re$V$, \,\, Im$V$),
$$ V=\frac{1}{G}\,e^{i\delta}\, \frac{\partial\bar
W}{\partial\bar\Phi}\,, $$
where $G=\partial^2 K/\partial\Phi \partial\bar{\Phi}$;

(ii) These curves do not touch the points where the vector field
$V$ vanishes (critical points of $V$).

Then, let us take such a closed curve $\Gamma_s$,
and pick up some parametrization $y$ on it,
so that the points on $\Gamma_s$
have coordinates
$\Phi_s(y),\,\, 0<y<1 $. Now, we want to find such
a change of parametrization $y(x)$ that
\begin{equation} \label{rep}
\Phi_s(y(x))\equiv \Phi(x,s)
\end{equation}
solves Eq. (\ref{bpseq}).
Substituting (\ref{rep}) into (\ref{bpseq})
we get
\begin{equation} \label{repe}
\frac{dx(y)}{dy}=G(\Phi_s(y), \bar{\Phi}_s(y))
\frac{d\Phi_s}{dy} e^{-i\delta}
\left(\left.
\frac{d\bar{W}}{d\bar{\Phi}}\right|_{\Phi=\Phi_s(y)}\right)^{-1}\,.
\end{equation}
The right-hand side of (\ref{repe}) is
defined and positive
if and only if the curve corresponding to $\Phi_s$ is tangent
to the vector field $V$ and does not touch its critical points
(where the right-hand side of (\ref{repe}) goes to
$+ \infty$).
In this case we can integrate (\ref{repe}) to get a
monotonous function $x(y)$.
The function $l(s)$ is given by the integral of
the right-hand side from 0 to 1,
\begin{equation}
l(s)=\int_{0}^{1} dy G(\Phi_s(y), \bar{\Phi}_s(y))
\frac{d\Phi_s}{dy} e^{-i\delta}
\left(\left.
\frac{d\bar{W}}{d\bar{\Phi}}\right|_{\Phi=\Phi_s(y)}\right)^{-1}\,.
\label{repl}
\end{equation}
Here, we set $y(0)=0,\,y(l(s))=1$.
Note, that the space ${\cal S}$ does not depend on
the nondegenerate changes of metric -- such changes
only affect the function $l(s)$.

Now, let us discuss the structure of ${\cal S}$
in more detail.
The following fact is very helpful --
the vector field $V$ is orthogonal to the gradient of the
multivalued function $I(\Phi,\bar{\Phi})$,
\begin{equation} 
I =\frac{e^{-i\delta} W[\Phi (x)] - e^{i\delta} \bar W[\bar
\Phi (x)]}{2i}\,,
\label{integral}
\end{equation}
which is actually the integral of motion
of Eq. (\ref{bpseq}),
$$
\dot{I} = \frac{1}{2i}\left( \frac{\partial W}{\partial\Phi }\, 
e^{-i\delta}\,
\dot{\Phi} - \frac{\partial \bar W}{\partial\bar\Phi}\, e^{i\delta}\,
\dot{\bar\Phi} \right)= 0\, .
$$
(Note, that expression (\ref{integral}) is the integral of
motion  in the multidimensional problem, with the arbitrary number
of fields $\Phi^i$, $\bar\Phi^{\bar j}$.)

Therefore, the closed curves tangent to $V$ present a net of  the 
level lines of $I$,
i.e. the curves $\Gamma_s$ are described
by the equation
\begin{equation}
I (\Phi_s)=s\,.
\label{is}
\end{equation}
At the same time, for the given $s$, the space of solutions of Eq.
(\ref{is})
could have several components, and
not all of them would correspond to
closed curves -- some of the components could even be noncompact
(see Sec. 3 for more details).

Now, we can describe ${\cal S}$ as the space of pairs $(s,a)$
where $s$ is such that Eq. (\ref{is}) contains
at least one component that is a closed curve,
and the index $a$ just numerates such components.
From the description above we conclude that the space ${\cal S}$
is an open space.

In fact, for a pair $(s,a)$, and the corresponding curve
$\Phi_{s,a}(y)$, we construct
its deformation $(s+\Delta s)$, $\Delta \Phi_{s,a}$,
\begin{equation} \label{def}
\frac{\partial I}{\partial \Phi} \Delta \Phi_{s,a} +
\frac{\partial I}{\partial \bar{\Phi}} \Delta \bar{\Phi}_{s,a} =
\Delta s\,.
\end{equation}
Equation (\ref{def}) can always be solved
(for small $\Delta s$) because the closed curves do not touch
the space of zeroes of the gradient of $I$ (that coincides with
the space of zeros of $dW$).
Therefore, the allowed values of $s$ is a union of finite intervals
$(s_k, s_{k+1})$,
and, possibly, semi-infinite intervals
$(-\infty, s_i) $ or $(s_f, +\infty)$. 
It is also possible that the allowed values of $s$
form the full line $(-\infty, \infty)$.

To complete the description of the space ${\cal S}$ we
should understand what happens with the closed curve
when $s$ reaches its critical values.
If $M$ is compact, all critical values
$s_k$ have to be finite, and
the critical curve $\Phi_{s_k,a}(y)$ is a
curve passing through one of the critical points
of $dW$.

In fact, as one can see from Eq. (\ref{repl}),
when the curve approaches a critical point of $dW$,
the derivative $d\bar{W}/d\bar{\Phi}$ in the denominator
vanishes, so the integrand (and, thus, the integral $l(s)$) tends
to $+\infty$.

If $M$ is noncompact,
we can think of it as of the space $\bar{M}$
with several
points $P_{\alpha}$ deleted.
Now, we will analytically continue $dW$ to the points
$P_{\alpha}$ keeping in mind examining the
behavior in these points.

If $dW$ could be analytically continued to the point
$P_{\alpha}$,
we could just add all such points to the space $M$
to get the space
$M_1$. Suppose that there is a closed curve close
to the point $P_{\alpha}$ -- then we expect
that there is also a critical closed curve, with
the critical value $s_c$
passing through the point $P_{\alpha}$.
We expect that $l(s)$ goes to infinity
as $s$ tends to $s_c$. From the standpoint
of $M$, the critical curve starts at
infinity and ends there. (If $l(s_c)$ is finite this means that
the manifold $M$ is geodesically noncomplete;
we discard  this possibility as an obvious pathology.)

Another case to be considered
corresponds to $dW$ having a
pole at $P_{\alpha}$.
Then, when $s$ tends to one of the infinities,
the whole curve $\Phi_{s,a}$ runs away towards
the point
$P_{\alpha}$. If the metric $G$ can be smoothly
continued to $P_{\alpha}$ then
$l(s)$ tends to zero as $s$ tends to
the corresponding infinity.
(Note, that if the metric  $G$ is also singular at the point
$P_{\alpha}$, we have a competition, but we will not discuss
such cases of competition here.)

After  the structure of ${\cal S}$ is determined, we complete
the first step
of our search for the classical BPS configurations.
The second step is to solve Eq. (\ref{lsl}) on ${\cal S}$.
Note, that the function
$l(s)$ depends on the
choice of metric $G$, which can be taken
to be an arbitrary smooth nonvanishing function.
Therefore, the behavior of
$l(s)$ inside the interval $(s_k,s_{k+1})$ is absolutely nonuniversal
and can be
chosen at will with the appropriate choice of $G$.
Nevertheless, we understand the asymptotics
of $l(s)$ when $s$ tends to its critical values.

From the analysis above we see
that if the critical value is finite,
$l(s)$ tends to $+\infty$.
If the critical value is infinite
and the scalar potential tends to infinity,
then $l(s)$ tends to zero.

This knowledge is sufficient to provide some information about the 
space 
of solutions of Eq. (\ref{lsl}) which determines the classical
BPS configurations.
This equation has {\it even} number of solutions
on finite intervals,
{\it odd} number of solutions
on semi-infinite intervals $(-\infty, s_k)$
and $(s_m, +\infty)$,
and {\it even} number of solutions in the special case when the
interval is $(-\infty, +\infty)$.
Note, that if the scalar potential does not grow
at infinities, we do not know the number of
solutions even modulo 2.

\subsection{Getting the Quantum BPS States from the Classical BPS
Solutions}
The quantum BPS solitons of
the two-dimensional theory correspond to such one-particle states
that are annihilated by half of the supercharges.
It will be assumed that the soliton particle is at rest,
i.e. the states in the quantum theory
with the zero spatial momentum will be considered.

The BPS particle states are annihilated by
half of the
supercharges, and thus, form doublet
representations (short multiplets) of the supersymmetry 
algebra~\cite{fi}.
The fact the supermultiplet of the BPS solitons
contains two states can be readily seen within the
quasiclassical quantization.
A regular representation of ${\cal N}=2$ superalgebra in
1+1 dimensions is quadruplet -- it contains two bosonic and two 
fermionic states.
This is seen, for instance, from the inspection of the chiral superfield
$\Phi (x_L,\theta )$. This can also be directly inferred from the 
analysis
of non-BPS solitons in the quasiclassical approximation. Indeed, such 
soliton
is characterized by the following collective coordinates:
the soliton center $x_0$ and four (complex) fermion collective 
coordinates
$\eta_{1,2}$ and $\bar\eta_{1,2}$, reflecting the nontrivial action of 
all four
supercharges when applied to the bosonic solution. Upon 
quantization,
the collective coordinates are to be treated as (adiabatically) varying
functions of time $x_0(t),\,\, \eta_{1,2}(t),\,\, \bar\eta_{1,2}(t)$.  
The
quantum-mechanical (first quantized) Lagrangian takes the form
\begin{equation}
{\cal L} = m\dot x_0^2 + i m\bar\eta_j\dot \eta_j\,, \qquad j=1,2\, 
.
\label{qmlfq}
\end{equation}
where $m$ is the soliton mass, and the dot stands for the time 
derivative.  If
$\eta_{1,2}$ are the canonic coordinates, then $i\bar\eta_{1,2}$ are 
the
conjugate canonic momenta, which determines the commutation 
relations
\begin{equation}
\{\eta_i(t),\bar \eta_j(t)\} = \delta_{ij}\, .
\label{cqm}
\end{equation}
The latter have a matrix representation in terms of four-by-four 
matrices,
the Hamiltanian is a four-by-four matrix,
i.e. the dimension of the supermultiplet is four.
What changes upon the transition to the BPS saturated soliton?
Two out of four supercharges annihilate the soliton. Correspondingly,
only two supercharges act nontrivially, and there are
two fermionic collective coordinates, $\eta$ and
$\bar\eta$. Again $\eta$ is to be treated as the
canonic coordinate, $i\bar\eta$ is the conjugate momentum. The 
commutation
relation which ensues is realized in terms of two-by-two matrices 
($\eta 
\sim
\sigma_-\,,\,\,\,
\bar\eta \sim \sigma_+$). The dimension of the supermultiplet is 
two.
This is the so-called shortened multiplet consisting of one bosonic 
and 
one
fermionic soliton. The very existence of the shortened multiplets is 
due 
to the
central extension of the original superalgebra, $Z\neq 0$.

It is well-known that 
quantization around the classical BPS configurations
results in the BPS state in the {\it perturbative}
quantum theory. In other words, every classical solution 
$\Phi(x,s_{*})$, where
$s_{*}$ is determined from Eq. (\ref{lsl}), gives rise to a BPS
quantum state order by
order in perturbation theory. This is so because of the multiplet 
shortening.
Naively, assuming that nonperturbative corrections
to the perturbative quantization do
not change this, we get quite a weird
picture of the BPS states in the model under consideration.
Their number is arbitrary, depending on details of
the metric $G$ -- even small variations of the metric could lead to
the appearance of new states or disappearance of the previously 
existing 
ones
(see Sec. 3 for details).

Moreover, there is an analog of Witten's index~\cite{Witten:1982df} 
that
counts the number of doublets in the centrally extended
${\cal N}=2$ superalgebra --
the Cecotti--Fendley--Intriligator--Vafa index.
This index is known to be independent of
the metric (see Appendix), which contradicts  our naive conclusion, 
based on
the perturbative quasiclassical quantization, that
the number of  the quantum BPS states equals the number
of the classical BPS solutions.

What saves the day is the observation that the number of the
classical BPS configuration modulo 2
is independent of the metric. This follows from the analysis in the 
previous
subsection.
Thus, we have to conclude that nonperturbative quantum effects
(like instantons in Witten's quantum mechanics) lift the BPS 
saturation 
of the
classical solutions and change
the structure of the representations.
Shortened multiplets pair up; a pair of the doublet
representations can form a regular quadruplet (non-BPS) 
representation.
The supercharges
which acted trivially at the perturbative level start
acting between the states that correspond to
the classical BPS configurations,
making quadruplets from pairs of doublets.
That is why the number of doublets in the
full {\em quantum theory}
equals the number of quasiclassical doublets
only mod 2. Pairing up of two classical BPS solutions
giving rise to a non-BPS quadruplet of quantum states
can occur at strong coupling.

The mass shift from the BPS bound is determined by the action of
the field configuration $\Phi (t,x)$ which smoothly interpolates
between the two given classical BPS solutions $\Phi_1 (x)$ and 
$\Phi_2 (x)$,
which pair up together.
With the  appropriate choice
of the metric, the barrier in the space of fields separating  the
two classical solutions, can be made  high. Correspondingly,
although the BPS saturation will be lifted at the (nonperturbative)
quantum level, the mass of the quadruplet representation will
be different from the BPS bound (i.e. from the central charge)
only  exponentially. We should also note that
in the problem of the domain walls (as opposed to 
solitons in 1+1 dimensions), the action is  proportional, in addition, 
to the wall
area. Therefore, the classical BPS wall with the infinite 
area remains BPS at the quantum level too, there is no tunneling.

Finally, we note that
using the independence of the CFIV index on the metric,
we can show that the finite intervals do not contribute to the CFIV
index, while each semi-infinite interval gives contribution equal to 1.
In fact, rescaling $G \rightarrow \lambda
G$, we change the
function $l(s)$ to $\lambda l(s)$.
Let $l_0$ be the minimum
 of $l(s)$ on the finite interval. Being
an integral over the positive integrand, $l_0>0$ .
 If $\lambda$ is
 larger than $L/l_{0}$, there are no
 classical BPS configurations coming from the finite interval.
  If we do the
same on the semi-infinite interval, then we will still get the 
contribution
from the region around infinity, where $l(s)$ tends to zero,
and this contribution is always 1.

\subsection{Multidimensional Generalization}

The consideration above can be generalized to the
multidimensional case as follows.
If the manifold $M$ is compact, then
the space ${\cal S}$ of the periodic trajectories of
Eq. (\ref{bpseq}) is bound by trajectories
that pass through the critical points of $dW$.
Thus, the function $l(s)$, being positive inside ${\cal S}$
and tending to $+\infty$ at the boundaries, has a
nonzero minimum $l_0$, and, as in the previous subsection, we can 
get 
rid of the classical
BPS configurations
by rescaling the K\"{a}hler metric.  Therefore, due to the CFIV index 
argument,
there will be no quantum BPS states.

Now, suppose that $M$ is noncompact,
and it is obtained from the compact manifold
$\bar{M}$ by cutting out several submanifolds
$N_{\alpha}$ of complex codimension 1.
For the sake of simplicity,
in this paper we will restrict ourselves to the case where
 submanifolds
$N_{\alpha}$ are smooth and do not intersect with each other,
\begin{equation}
M= \bar{M}  \setminus \cup N_{\alpha}\,.
\end{equation}
Suppose that the 1-differential $dW$  has
simple poles on these submanifolds,
\begin{equation}
\int_{\Gamma_{\alpha}} dW = \Delta_{\alpha} W\,,
\end{equation}
where $\Gamma_{\alpha}$ is a cycle in the vicinity of
$N_{\alpha}$ that cannot
be contracted to a point without
crossing $N_{\alpha}$, i.e. it has a nontrivial
linking with $N_{\alpha}$.

We claim that the number of the BPS states coming from
the quantization of the space $C_{\alpha}$ of the parametrized
curves $\Phi^i(x)$ that could be deformed to
$\Gamma_{\alpha}$ is bound from below by the Euler
number of $N_{\alpha}$.

Let us study the space $S_{\alpha}$ of periodic trajectories
in the space $C_{\alpha}$. Rescaling the K\"{a}hler metric
$$
G_{i\bar{j}} \rightarrow
\lambda G_{i \bar{j}}
$$
and taking $\lambda$ to $+\infty$ forces  $l(s)$ to
go to $+\infty$ everywhere except for the curves in the vicinity
of $N_{\alpha}$. Thus,
the problem is reduced to the vicinity of $N_{\alpha}$
that looks
as $C^* \times N_{\alpha} $, and
the differential is $c d\Phi/\Phi$, where
$\Phi$ is  the coordinate on $C^*=C \setminus 0 $.
Let us take the metric $G$ in the vicinity of $N_{\alpha}$
in the following form:
 \begin{equation}
G^{i\bar{j}}=
G_{\rm pr}^{i \bar{j}}
+t g^{i \bar{j}}
 + O(t^2)\,,
\end{equation}
where
$G_{\rm pr}^{i \bar{j}}$ is the inverse product metric,
  $g^{i \bar{j}}=0$ if $i>1, \bar{j}>1$ or $ i=\bar{j}=1$;
$g^{i \bar{1}}=v^i $ and $v^i$ is a vector field on $N_{\alpha}$.
For small $t$ the trajectory
starting at the point $(\Phi,P)$, whose projection on $C^*$
is periodic, will be nonperiodic in the $N_{\alpha}$ direction.
The shift in this direction is proportional to $v^i(P)$, i.e.
to the value of the vector field $v^i$ at the point $P$ on $N_{\alpha}$.
Thus, the number of the periodic trajectories is given by the number
of the zeroes of the vector field $v^i$, i.e. it
is bound from below by the Euler number of $N_{\alpha}$.

\section{ A Typical Example}
In this section we will consider a typical example that contains
most
of various cases considered in Sec. 2.
Consider
\begin{equation}
K(\Phi,\bar{\Phi})= \Phi\bar{\Phi}, \quad
dW=\frac{4\pi}{2 -
\cos\Phi}\, d\Phi   \,,
 \label{b1}
\end{equation}
The target space has the topology of  a
cylinder with two points deleted (Fig. 1),
\begin{equation}
-\infty < {\rm Im} \Phi <\infty\,,\qquad -\pi \leq {\rm Re} \Phi\leq 
\pi 
\,,
\end{equation}
and
\begin{equation}
(\Phi_*)_{1,2} = \pm i\ln \left(2+\sqrt{3}
\right)\,.
\end{equation}
\begin{figure}[h]
\vskip3mm
\epsfxsize=8.5cm
\centerline{\epsfbox{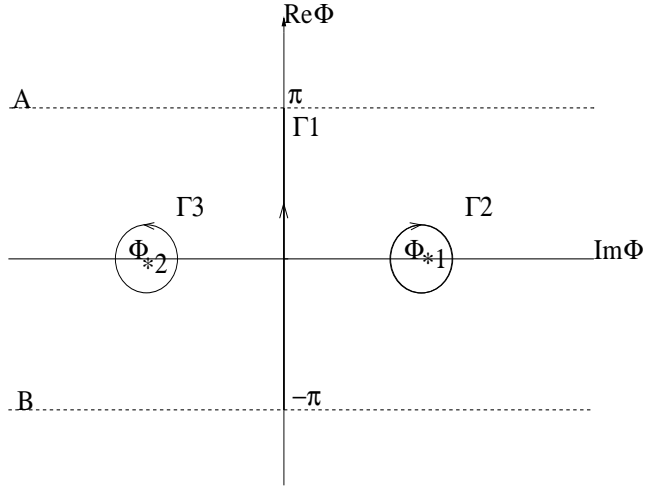}}
\caption{The target space in the problem (\ref{b1}). The dashed lines 
are glued, so
that the points A and B are actually one and the same point.}
\end{figure}

Correspondingly, there  are three noncontractible cycles, $\Gamma_1,
\Gamma_2$ and
$\Gamma_3$ in Fig. 1. The scalar potential is depicted in Fig. 2. Later 
on we will
deform the K\"ahler potential in Eq. (\ref{b1}) by adding a small 
perturbation.
 Since
$dW/d\Phi$ vanishes only at $|{\rm Im}\Phi|\rightarrow \infty$, the 
model has the
run-away vacua. The soliton  solutions stabilize the theory, as in
Ref.~\cite{Dvali:1999ht}. The periods corresponding to the
cycles $\Gamma_1, \Gamma_2$ and $\Gamma_3$ are  all equal to
\begin{equation}
\Delta W = \frac{8\pi^2}{\sqrt{3}} \,.
\label{per}
\end{equation}

\begin{figure}[h]
\vskip3mm
\epsfxsize=8.5cm
 \centerline{\epsfbox{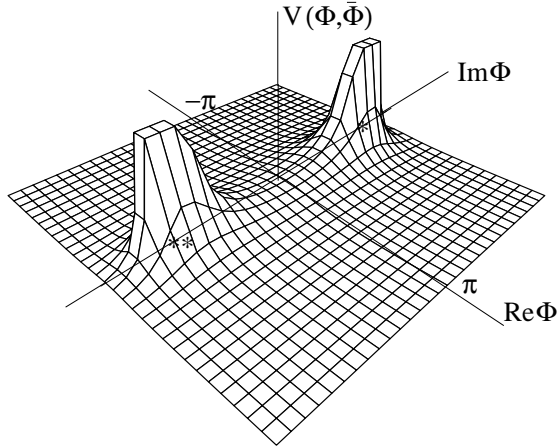}}
 \caption{The scalar potential $V(\Phi,\bar{\Phi})$ in the problem 
(\ref{b1}) near
 $(\Phi_{*})_{1,2}$. Here * denotes $\Phi_{*1}$, ** denotes 
$\Phi_{*2}$.}
\end{figure}

In the problems with one variable $\Phi$,
the fact of existence of the integral of motion (\ref{integral})
is extremely helpful. By inspecting Fig. 1, one immediately infers
that one can expect three solutions of  Eq. (\ref{bpseq}) --
one connecting the points $\Phi = -\pi$ and $\Phi = \pi$
along the real axis of $\Phi$, and two other solutions winding
around $(\Phi_{*})_{1,2}$.
Let us discuss them in turn.
\begin{figure}[h]
\vskip3mm
\epsfxsize=8.5cm
 \centerline{\epsfbox{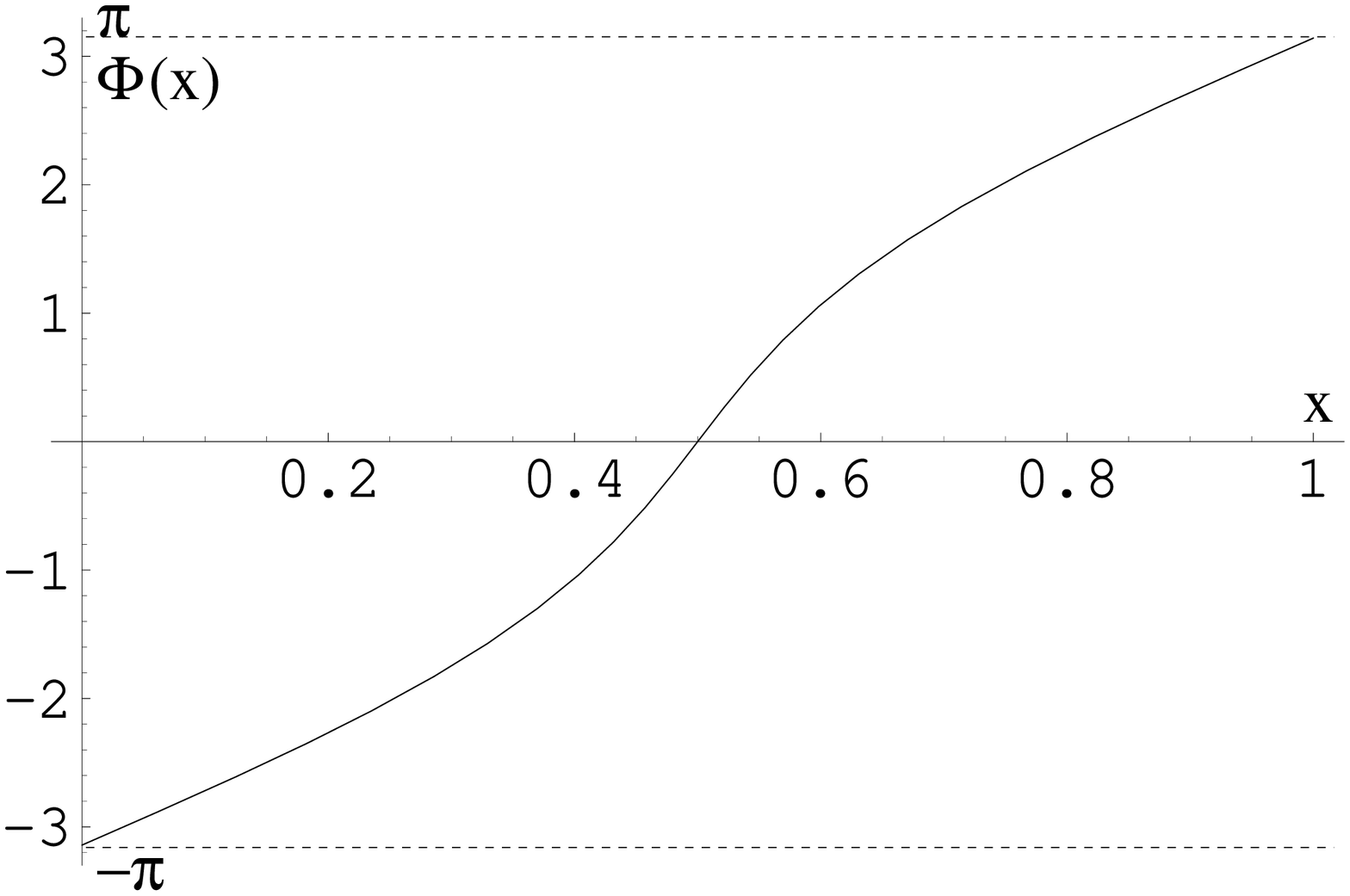}}
 \caption{$\Phi$ versus $x$ for the trajectory running along 
$\Gamma_1$.}
\end{figure}

For real $\Phi$ both equations in (\ref{bpseq}) coincide.
The solution $\Phi (x,s) $ is readily obtained in an implicit form. It 
is given by
inversion of the formula
\begin{equation}
x-\frac{1}{2}=\frac{\Phi}{2\pi}
-\frac{1}{4\pi} \sin\Phi\, .
 \label{b4}
\end{equation}
Here, $\delta=0$, which follows from Eq. (\ref{per}). And we omit the 
index $s$ since in this case $s=I=0$,
just a fixed number.
In the problem
at hand each trajectory is in one-to-one
correspondence with the value of $I$. So, we will label the 
trajectories
by the corresponding value of $I$ instead of $s$ in this section. We 
obviously
have $\l(I=0) = 1$, (as noted
before, we set $L=1$ here for convenience).
The function $\Phi (x) $ is depicted in Fig. 3, while
the energy density corresponding to this solution is plotted in Fig. 4.
Of course, the center of the soliton (1/2 in Eq. (\ref{b4}))
can be chosen arbitrarily.
As we will see shortly, this trajectory is exceptional.
\begin{figure}[h]
\vskip3mm
\epsfxsize=8.5cm
\centerline{\epsfbox{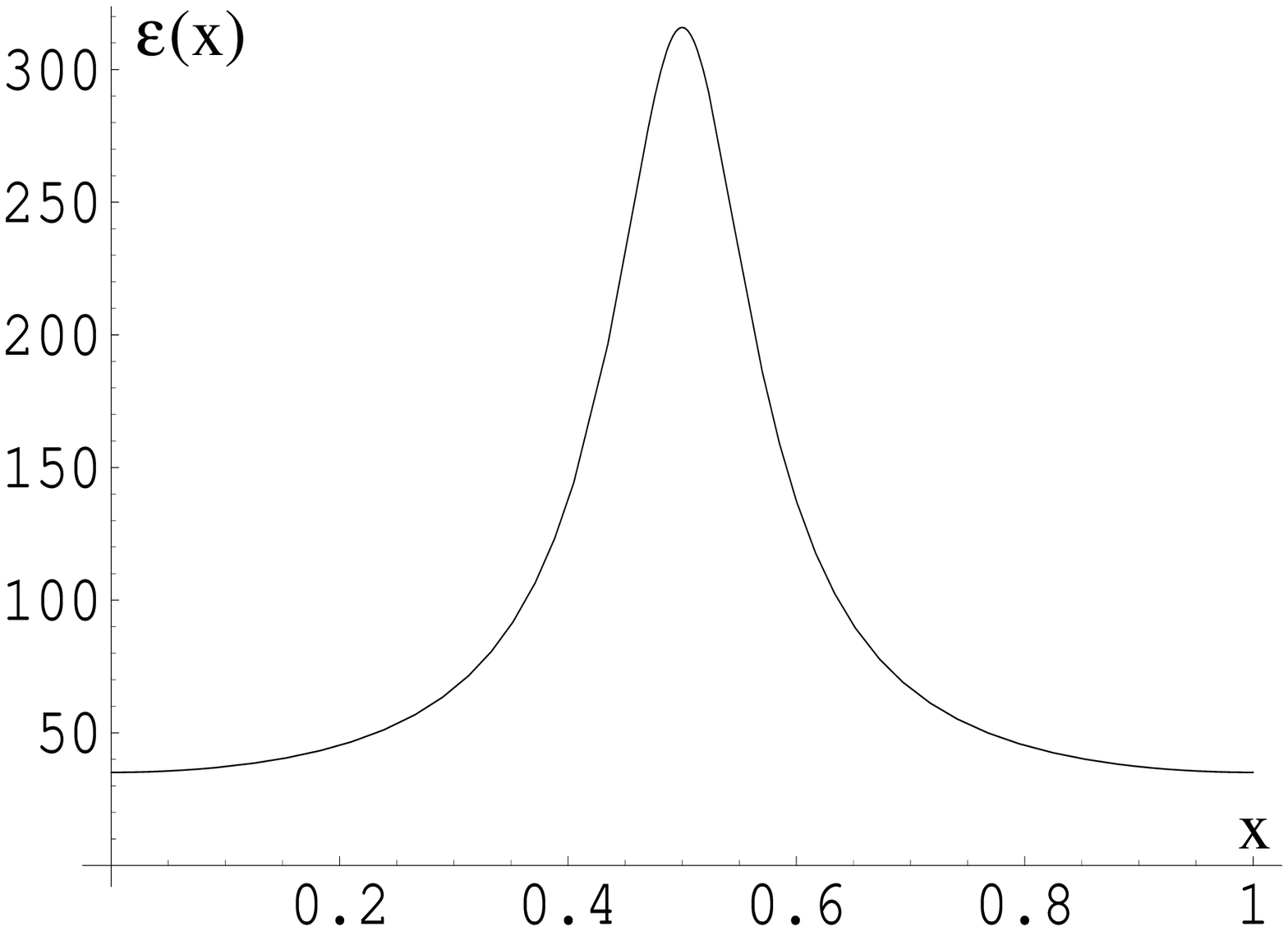}}
\caption{The energy density $\epsilon (x)=\dot\Phi (x)^2+V[\Phi 
(x)]$ 
versus $x$ for the $\Gamma_1$ soliton solution.}
\end{figure}

The solitons corresponding to $\Gamma_{2,3}$ can be established as 
follows.
We will focus on $\Gamma_{2}$ since the solution for $\Gamma_{3}$
is similar. The superpotential is obtained as
\begin{equation}
W = \frac{8\pi}{\sqrt{3}}\, {\rm arctan}
\left(\sqrt{3}\tan\frac{\Phi}{2}\right)\,.
 \label{b5}
\end{equation}
Here we have chosen a specific branch of the multivalued 
superpotential 
$W$.
Correspondingly, for the trajectories winding around
$(\Phi_*)_1$ the value of $I = {\rm Im}\, W$ spans the interval
$I =  (0,\infty)$.
Following (\ref{is}), we can get $\Phi(x,I)$ implicitly from
\begin{displaymath}
\mbox{Im} W(\Phi_{I})=I\,.
\end{displaymath}
Then, we can parametrize the corresponding closed curve 
$\Gamma_{I}$ as
$\Phi_{I}(y), 0<y<1$ as in Sec. 2. Then, from Eq. (\ref{ls}), we can get 
$l(I)$.
Let us write
$$
\Phi = \phi + i\chi
$$
where $\phi$ and $\chi$ are real functions of $x$. The condition
$$
I (\phi ,\chi ) = \,\mbox{a positive constant}
$$
defines a family of trajectories (see Sec. 2). Practically,
for each given trajectory
one can find analytically $\phi_I (x)$ and $\chi_I(x)$
using the BPS saturation equation
\begin{equation}
\dot\Phi =  \frac{4\pi}{2-\cos\bar{\Phi}}\, ,
\label{bpseqone}
\end{equation}
and then obtain $l(I)$. The period function $l(I)$ {\it versus} $I$ is 
shown in Fig. 6. The solution corresponding to the BPS soliton is 
obtained from 
the condition $l(I_0) = 1$. The energy density corresponding to the 
$\Gamma_2$ 
soliton solution is shown in Fig. 5. Note that we denote the class of 
the 
trajectories homotopical to
the $\Gamma_1$
 cycle as the class $T_1$, such as the trajectories $\Gamma^{'}, 
\Gamma^{''}$ in
 Fig. 7 below, and the classes of the trajectories homotopical to the
 $\Gamma_2, \Gamma_3$ cycles as the class $T_2, T_3$. And we 
know that
the  class
 $T_3$ could be treated absolutely in the same way as the class 
$T_2$.
\begin{figure}[h]
\vskip3mm
\epsfxsize=8.5cm
 \centerline{\epsfbox{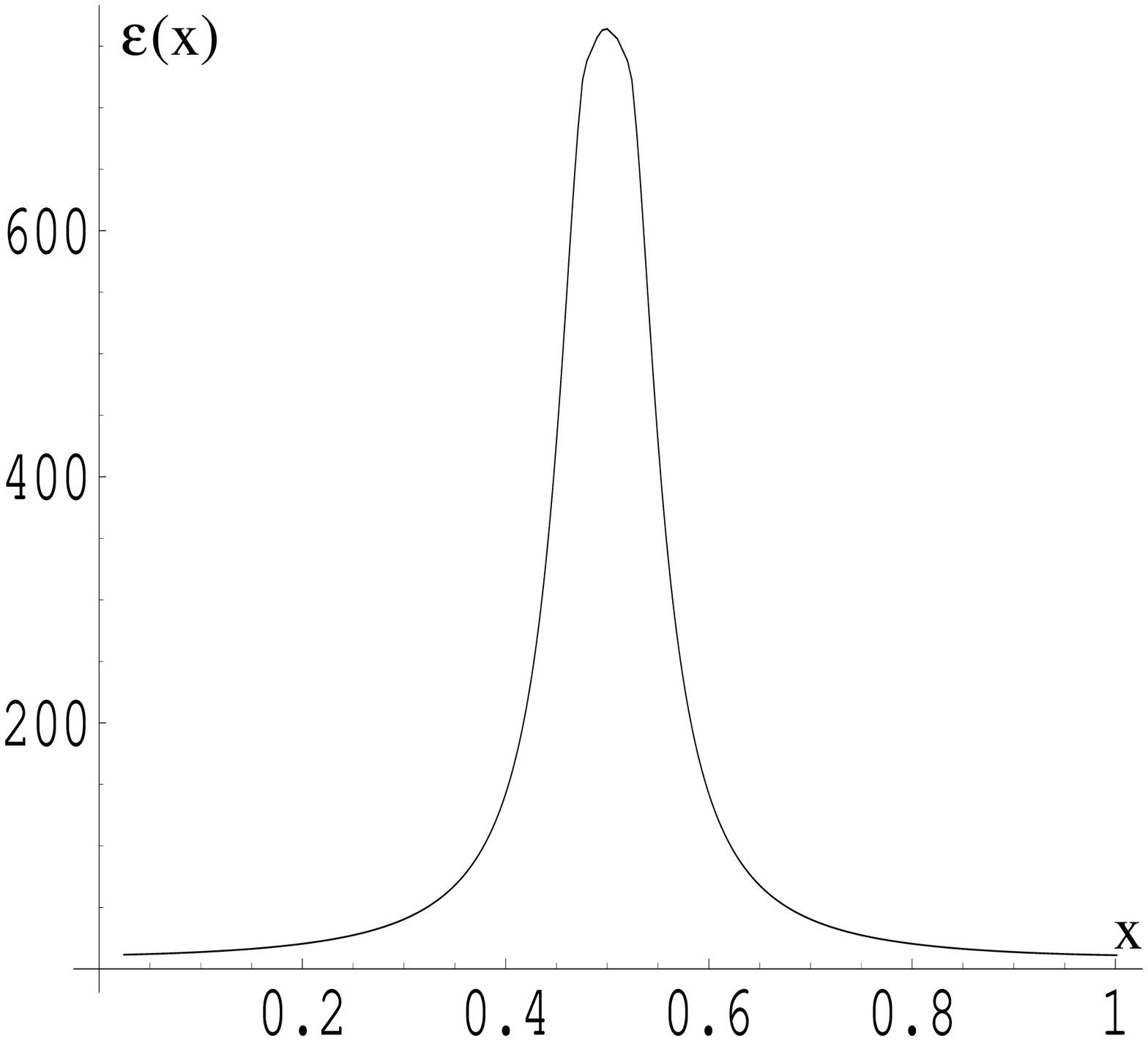}}
 \caption{The energy density $\epsilon (x)$ versus $x$ for the 
$\Gamma_2$
 soliton solution.}
\end{figure}

\begin{figure}
\vskip3mm
\epsfxsize=8.5cm
\centerline{\epsfbox{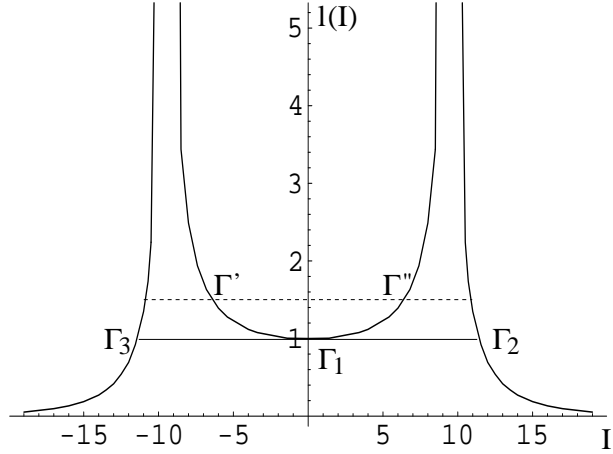}}
\caption{The period $l(I)$ versus $I$.}
\end{figure}
\begin{figure}[h]
\vskip3mm
\epsfxsize=8.5cm
\centerline{\epsfbox{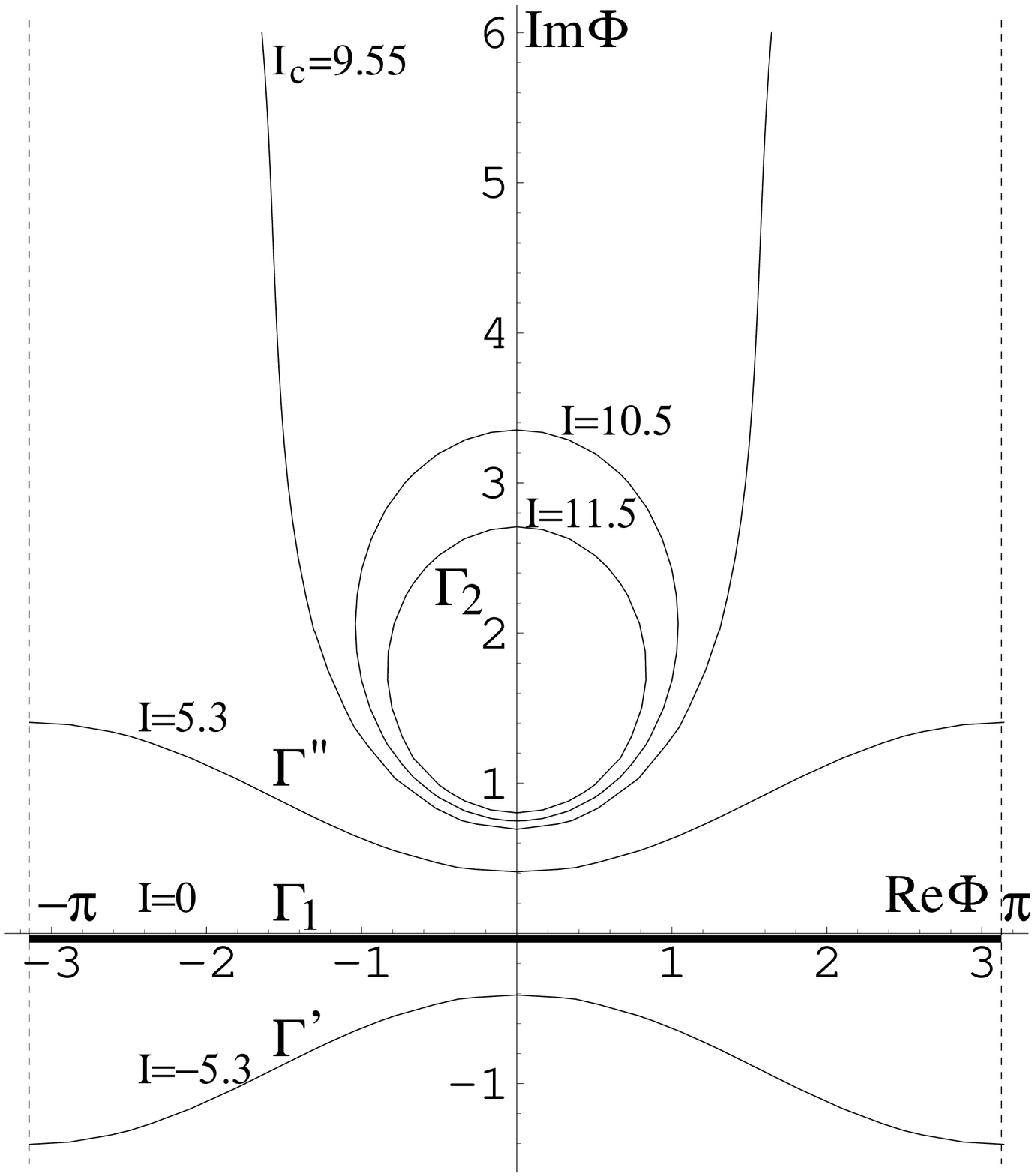}}
\caption{The typical trajectories for the homotopy classes $T_1$ and 
$T_2$.
 Here, $I_{c}$ is the corresponding value of $I$ for which the period 
$l(I)$ goes to infinity. And the dashed lines are glued.}
\end{figure}
Thus, classically we have three BPS soliton solutions preserving
one half of supersymmetry.
Let us see what happens in the weak coupling regime.
To this end we introduce a small coupling constant $g^2$ in the 
K\"ahler 
metric,
\begin{equation}
K(\Phi, \bar\Phi ) = \frac{1}{g^2}\Phi\bar\Phi\, .
\label{newk}
\end{equation}
One can view the factor $ {1}/{g^2}$ as a deformed metric.
The equation of the BPS saturation remains the same as in Eq. 
(\ref{bpseqone}),
if, instead of the original variable $x$, one introduces $\tilde x =
g^2 x$. The dot will now denote differentiation over $\tilde x$.
The solution is of the same type  as discussed above with the 
constraint 
that
the period in $\tilde x$ is $g^2\ll1$ rather than unity.
The BPS saturation equations have  no solution with the small period 
for 
 the
trajectory along
$\Gamma_1$ . (We hasten to note, however, that a nonsaturated 
solution 
of the
classical equations of motion,  winding around $\Gamma_1$ with the 
arbitrary
period, always exists, as it is perfectly clear from
the examination of  Fig. 1.)

The BPS saturation equations do have small period solutions   for the
$\Gamma_{2,3}$ cycles. The requirement $l(I)= g^2 \to 0$ implies 
that
the trajectories wind around $(\Phi_*)_{1,2}$ very close to
$(\Phi_*)_{1,2}$ (see Fig. 6 and 7). Then, say for the $\Gamma_2$
cycle, $dW/d\Phi$ can be
replaced by
\begin{equation}
\frac{dW}{d\Phi} = \frac{4\pi}{\sqrt{3}\, i}\, \frac{1}{\Phi - 
(\Phi_*)_{1}}\,,
\label{clsol}
\end{equation}
and the solution with the given $l$ takes the form
\begin{equation}
\Phi = (\Phi_*)_{1} + \left(\frac{2l}{\sqrt{3}}
\right)^{1/2}\, \exp \left(\frac{2\pi i\tilde x}{l}\right)\, ,\qquad l 
= g^2\,.
\label{clsom}
\end{equation}
This is precisely the solution which was first obtained in
Ref.~\cite{Dvali:1999ht}.

As was mentioned, the existence of the solution
preserving one half of SUSY at small coupling immediately translates 
at 
the quantum level into the presence in the spectrum of the BPS 
multiplet of 
particles.
That such particles neither appear nor disappear in the
process of evolution from small $g^2$ to $g^2=1$ is guaranteed by the
CFIV index. At this point, one may think that there are three BPS 
solutions at
the classical level and two BPS solutions at the quantum level in this 
example.
But it was said in Sec. 2.3 that the number of doublets in the 
quantum 
theory
equals the number of quasiclassical doublets only mod 2, which, at 
first 
sight, contradicts
this example. However, one should notice that the classical BPS 
solution
corresponding to the cycle $\Gamma_1$ presents actually a 
degeneration 
of two solutions that are ``glued" together, which can be seen clearly 
from 
Fig. 6
that the two solutions $\Gamma^{'}, \Gamma^{''}$ are ``glued" 
together 
at the point
corresponding to the $\Gamma_1$ solution, so that the relative 
fermion
charges of these two solutions are arranged in such a way
that their contributions to the CFIV index cancel each other. 
Therefore, 
this example is indeed in agreement with the analysis of Sec. 2.3.

Summarizing, in the given sample problem, with three 
noncontractible
cycles in the target space, we found three distinct soliton 
supermultiplets.
Two of them, corresponding to the trajectories winding around
the poles of $dW/d\Phi$, are BPS saturated (short multiplets,
one half of supersymmetry is preserved). The soliton
corresponding to the
$\Gamma_1$ cycle is not saturated, in spite of the fact that 
classically
one can find a solution of the BPS saturation equations in strong 
coupling
at certain (isolated) values of $g^2$. The classical solution is not 
elevated to the quantum level.

\section{The Quantum Mechanical Explanation
of Instantonic Corrections to the Quasiclassical BPS States}

Consideration of Sec. 2 may leave a wrong impression
that the phenomenon of localization to the poles of $dW$
in the computation of the number of the BPS doublets is quite
specific to $d=2$ theories.  In this section we will show
that (as it always happens with phenomena in the computation of
various
index-like quantities) this phenomenon
is a special case
of a more general phenomenon that occurs
in ${\cal N}=2$ quantum mechanics with the target space
with isometry (the Killing vector).
Moreover, we will show that the mysterious nonperturbative
corrections mentioned in Sec. 2 could be easily understood
as instantonic corrections in Witten's ${\cal N}=1$ quantum 
mechanics
associated with the ${\cal N}=2$ quantum mechanics on the
target space with isometry.

 \subsection{Doublets of ${\cal N}=2$ $(d=2)$ Superalgebra and 
Ground
States of Associated Quantum Mechanics}
 The algebra of supersymmetries of
the ${\cal N}=2$ two-dimensional field theory has the following form:
 \begin{equation}
\label{alg} \{ Q_{\pm}, \bar{Q}_{\pm}  \}= H \pm P \,,\,\,
\{ Q_{+}, Q_{-}  \}= Z \,,\,\,
\{ \bar{Q}_{+}, \bar{Q}_{-}  \}= \bar{Z}\,. \label{sal}
\end{equation}
Here $P$ is the
momentum operator in the $x$ direction, $H$ is the Hamiltonian, the
operators
$Z$ and $\bar{Z}$
are complex conjugate to each other and are called the central
charges.  The
subscripts $+$ and $-$ of the supercharges denote that they
have the charges
$+1/2$ and $-1/2$ under the $SO(1,1)$ Lorentz group.  Operators
without the bar
are the chiral supercharges, while those with the bar antichiral
ones.

To simplify the study of the representations of the algebra 
(\ref{sal}), 
we
will make the following redefinition:
\begin{equation}
q_{1}=Q_{+},\,\,
\bar{q}_{1}=\bar{Q}_{+},\,\,
q_{2}=\bar{Q}_{-},\,\,
\bar{q}_{2}=Q_{-}\,.
\end{equation}

Then the commutation relations (\ref{alg}) take a well-known  form 
of 
the Clifford algebra
\begin{equation}  \label{alq}
\{ q_{\alpha} , \bar{q}_{\bar{\beta}} \} = N_{\alpha \bar{\beta}}\,,
\end{equation}
where $\alpha,\bar{\beta}=1,2$ and the $2 \times 2 $ matrix $N$ is
\begin{equation} \label{nhpz}
N =   \left[ \begin{array}{cc}
H+P & Z \\ \bar{Z}& H-P
\end{array} \right]\,.
\end{equation}
We are interested in the states that represent particles at rest,
so we will restrict ourselves to the representations with $P=0$.

In the irreducible representations of the algebra (\ref{alq}),
$H$ and $Z$ are represented by numbers $E$ and $Z$. We are
interested
in such representations that $Z \neq 0$.
Then the irreducible representation is either 4-dimensional,
if the matrix $N$ is nondegenerate, or two dimensional,
if the matrix $N$ has a zero eigenvalue.
The latter happens if and only if
\begin{equation}
Z=e^{i\theta}E\,,
\end{equation}
where $\theta$ is a real constant.
The shortening of the irreducible representation
could also be interpreted as follows.
The ${\cal N}=2$ algebra contains, as a subalgebra, the
following algebra of an associated ${\cal N}=1$ quantum mechanics:
\begin{equation}  \label{qm}
\{ q_{\theta} , \bar{q}_{\theta} \}=H_{\theta}\,,
\end{equation}
where
\begin{equation} \label{qdel}
q_{\theta} = q_{1} + e^{i\theta} q_{2},\,\,
\bar{q}_{\theta}= \bar{q}_{1}+e^{-i\theta}\bar{q}_{2},\,\,
H_{\theta}=2H+e^{-i\theta}Z+e^{i\theta}\bar{Z}\,.
\end{equation}

Then, the doublet representations of the algebra
(\ref{alq}) are in one-to-one correspondence
with the ground states of the associated ${\cal N}=1$ quantum 
mechanics
(\ref{qm}).

\subsection{The Representation of the ${\cal N}=2$ ($d=2$) Algebra
 from the
Representation of the ${\cal N}=2$ Quantum Mechanics on the Target 
Space
with an Isometry}

Let us start from the conventional superalgebra of ${\cal N}=2$ 
quantum mechanics

\begin{equation} \label{oqm}
\{ q_{1,0}, \bar{q}_{1,0} \}=
\{ q_{2,0}, \bar{q}_{2,0} \}=H\,.
\end{equation}

The ${\cal N}=2$ quantum mechanics, obtained by dimensional 
reduction
of the generalized Landau-Ginzburg theory, provides the following
representation of the algebra (\ref{oqm}).

Consider the Clifford algebra
\begin{equation}
\{ \psi_{+}^{i} , \bar{\psi}_{+}^{\bar{j}}   \} = G^{i \bar{j}};\quad
\{ \psi_{-}^{i} , \bar{\psi}_{-}^{\bar{j}}   \} = G^{i \bar{j}}.
\end{equation}
Then, the algebra (\ref{oqm}) has the following representation:
\begin{eqnarray} \label{repqm}
 Q_{+,0} & \!\!=&\!\!  q_{1,0}  =
  \bar{\psi}_{-}^{\bar{j}}
 \frac{\partial}{\partial \bar{\Phi}^{\bar{j}}}
 +\psi_{+}^{j} \Omega_j
 -\psi^{l}_{+}\bar{\psi}^{\bar{i}}_{-
}\bar{\psi}^{\bar{j}}_{+}\frac{\partial
 G_{l\bar{j}}}{\partial \bar{\Phi}^{\bar{i}}}\,
 ,\nonumber \\[0.2cm]
 Q_{-,0} & \!\!=&\!\!  \bar{q}_{2,0}  = \bar{\psi}_{+}^{\bar{j}}
 \frac{\partial}{\partial \bar{\Phi}^{\bar{j}}} + \psi_{-}^{j}
\Omega_j 
-\psi^{l}_{-}\bar{\psi}^{\bar{i}}_{-
}\bar{\psi}^{\bar{j}}_{+}\frac{\partial
 G_{l\bar{j}}}{\partial \bar{\Phi}^{\bar{i}}} \,
 ,\nonumber \\[0.2cm]
 \bar{Q}_{+,0} & \!\!=&\!\!  \bar{q}_{1,0}  = \psi_{-}^{j}
 \frac{\partial}{\partial \Phi^{j}} +\bar{\psi}_{+}^{\bar{j}}
 \bar{\Omega}_{\bar{j}}
 -\bar{\psi}^{\bar{l}}_{+}\psi^{i}_{-}\psi^{j}_{+}\frac{\partial
 G_{j\bar{l}}}{\partial\Phi^{i}}\,,\nonumber \\[0.2cm]
  \bar{Q}_{-,0} & \!\!=&\!\!
 q_{2,0}  = \psi_{+}^{j} \frac{\partial}{\partial \Phi^{j}}
 +\bar{\psi}_{-}^{\bar{j}} \bar{\Omega}_{\bar{j}}
 -\bar{\psi}^{\bar{l}}_{-}\psi^{i}_{-}\psi^{j}_{+}\frac{\partial
 G_{j\bar{l}}}{\partial\Phi^{i}}\,.
 \end{eqnarray}
Suppose that the target space has an isometry that
 preserves both the metric and the one-differential.
 It means that there is a real Killing vector field
\footnote{Here and below we will assume that
$m$ is a real index, i.e. it can take both holomorphic
($i$), and antiholomorphic ($\bar{i}$) values.}
 \begin{displaymath}
 v^m=(v^i,\bar{v}^{\bar{i}})\,,
 \end{displaymath}
 and the Lie derivative along this
 real vector field $L_v+ \bar{L}_{\bar{v}}$
 leaves both the metric $G$ and the differential $\Omega$
 invariant.
 Here $L_v$ acts on the $(k,l)$ tensor as follows:
 \begin{equation}
  L_v
  T_{i_1 \ldots i_k, \bar{j}_1, \ldots, \bar{j}_l}=
 v^j \frac{\partial}{\partial \Phi^j}
  T_{i_1 \ldots i_k, \bar{j}_1, \ldots, \bar{j}_l}+
 \sum_{a=1}^{k}
  T_{i_1 \ldots i_{a-1} i \ldots i_k, \bar{j}_1, \ldots, \bar{j}_l}
 \frac{\partial v^{i}}{\partial \Phi^{i_a}},
 \end{equation}
 and $\bar{L}_{\bar{v}}$ is complex conjugated to
 $L_v$.

 This invariance means that
\begin{equation}  \label{lig}
 \left(v^j \frac{\partial}{\partial \Phi^j}+
 \bar{v}^{\bar{j}} \frac{\partial}{\partial \bar{\Phi}^{\bar{j}}} 
\right)
 G_{i \bar{k}}+
 G_{j \bar{k}}
 \frac{\partial v^j}{\partial \Phi^i}-
 G_{i \bar{j}}
 \frac{\partial \bar{v}^{\bar{k}}}{\partial \bar{\Phi}^{\bar{j}}}=0\,,
\end{equation}

\begin{equation}  \label{lio}
 v^j \frac{\partial}{\partial \Phi^j} \Omega_{i}+
 \Omega_{j}\frac{\partial v^j}{\partial \Phi^i}=0\,.
\end{equation}
 Note, that the antiholomorphic derivatives are absent in (\ref{lio})
 due to holomorphy of $\Omega$.

 Using this vector field, we can modify the representation of
 ${\cal N}=2$ supersymmetric quantum mechanics in such a way that 
it
 represents the algebra of ${\cal N}=2$ ($d=2$) supersymmetry
in the following way:
\begin{eqnarray}
q_{1}=Q_{+} & = & Q_{+,0}+\frac{1}{2}
 G_{i \bar{j}} v^{i} \bar{\psi}_{-}^{\bar{j}}\,,\nonumber \\
\bar{q}_{2}=Q_{-} & = & Q_{-,0}+\frac{1}{2}
 G_{i \bar{j}} v^{i} \bar{\psi}_{+}^{\bar{j}}\,,\nonumber \\
\bar{q}_{1}=\bar{Q}_{+} & = & \bar{Q}_{+,0}+\frac{1}{2}
 G_{i \bar{j}} \bar{v}^{\bar{j}} \psi_{-}^{i}\,,\nonumber \\
q_{2}=\bar{Q}_{-} & = & \bar{Q}_{-,0}+\frac{1}{2}
 G_{i \bar{j}} \bar{v}^{\bar{j}} \psi_{+}^{i}\,.
\end{eqnarray}
Here
 $P$ is represented as
 \begin{equation}
 P \rightarrow (L_v+\bar{L}_{\bar{v}})\,,
 \end{equation}
 while the central term
\begin{equation}
 Z= v^i \Omega_i\,.
\end{equation}
Note that Eq. (\ref{lio}) implies that $Z$ is a constant.

 Now, let us have a closer look at the associated ${\cal N}=1$
 quantum mechanics.
 In particular, if we introduce
 $$
 \begin{array}{cccccc}
 \chi_{1}^{i}&=&e^{i\theta} \psi_{+}^{i},&
 \bar{\chi}_{1}^{\bar{i}} &=&\bar{\psi}_{-}^{\bar{i}}\\
 \chi_{2}^{i}&=&\psi_{-}^{i},&
 \bar{\chi}_{2}^{\bar{i}}&=&e^{-i\theta}\bar{\psi}_{+}^{\bar{i}}\,,
 \end{array}
 $$
then  the
 supercharge $q_{\theta}$ takes the form:
\begin{equation}
 q_{\theta} = \chi_{1}^{m} \left(\frac{\partial}{\partial \Phi^m} +
 \omega_m(\Phi,\bar{\Phi})\right)\,,
 \end{equation}
 where $\omega_m=(\omega_i, \bar{\omega}_{\bar{i}})$, and

 \begin{eqnarray}
 \omega_i&\!\!=&\!\!e^{-i\theta} \Omega_i +\frac{1}{2} G_{i
 \bar{j}}\bar{v}^{\bar{j}}-
 \bar{\psi}_{-}^{\bar{l}}\psi_{-}^{j}\frac{\partial 
G_{i\bar{l}}}{\partial
 \Phi^{j}} \,, \nonumber \\[0.2cm]
 \bar{\omega}_{\bar{i}}&\!\! =&\!\!
e^{\,\,i\theta}\bar{\Omega}_{\bar{i}}\,\,+\frac{1}{2}G_{i\bar{i}}v^{i}\,
-
\bar{\psi}_{+}^{\bar{j}}\psi_{+}^{l}\frac{\partial
G_{l\bar{j}}}{\partial \bar{\Phi}^{\bar{i}}}\,. \label{omega}
 \end{eqnarray}
 One can check that $\omega$ is closed,
 i.e.
 $$
 \frac{\partial \omega_m }{ \partial \Phi^{n}} =
 \frac{\partial \omega_n }{ \partial \Phi^{m}}\,.
 $$
 (Here, as above, $\Phi^m=(\Phi^i, \bar{\Phi}^{\bar{i}}$).)
 In the computation of Witten's index one can
 continuously change the
 superpotential (in a way preserving discretization of the spectrum).
 Therefore, we can study a family of
 the $\omega_i(\lambda)$ as the following
 \begin{equation}
 \omega_i(\lambda)=e^{-i\theta}\Omega_i +\frac{\lambda}{2}
\bar{v}^{\bar{j}}G_{i\bar{j}}-
\bar{\psi}_{-}^{\bar{l}}\psi_{-}^{j}\frac{\partial
G_{i\bar{l}}}{\partial \Phi^{j}}\,.
 \end{equation}
 As it is well known, the classical ground states correspond to zeroes
 of $\omega(\lambda)$.
 Let us tend
 $\lambda$ to $+\infty$.  At first, suppose that target space is
 compact. Then, the zeroes of $\omega(\lambda)$ tend to the zeroes 
of 
$v$,
 and if $v$ has no zeroes, Witten's index is equal to zero.  Now,
 suppose that the target space is not compact, and $\Omega$ has 
poles on
 its compactification.  Then, as $\lambda$ tends to $+\infty$, the
position of the zeroes of $\omega(\lambda)$ tends to the position of 
the
poles, and the computation of Witten's index is reduced to the
computation in the vicinity of the poles.

\subsection{${\cal N}=2$ ($d=2$) Field Theory as ${\cal N}=2$ 
Quantum Mechanics
 on the Loop Space}

The two-dimensional generalized Landau-Ginzburg model,
as the quantum theory in the given winding sector, is
an ${\cal N}=2$ quantum mechanics on the corresponding loop
space modified by the vector field that rotates the loop.
Various aspects of this phenomena were studied previously in
\cite{Cecotti, Nikita1, Nikita2, BLN}.

The coordinates on the loop space are
$$
 \Phi^{i,x}=\Phi^{i}(x)\,.
$$
The vector field $v^{i,x}$ generated by
 $x \rightarrow x+\epsilon$  is
\begin{equation}
 v^{i,x}=\frac{\partial \Phi^i(x)}{ \partial x}\,.
\end{equation}

Let us see how this modifies the supercharges.
For example, $Q_{+}$ takes the following form
(the sum over the continuous index $x$ is replaced by the integral)
\begin{equation}  \label{qqft}
 Q_{+}=
 \int dx \left\{ \bar{\psi}_{-}^{\bar{j}}(x)
 \left( \frac{\partial}{\partial \bar{\Phi}^{\bar{j}}}(x)+\frac{1}{2}
 G_{i\bar{j}}\frac{\partial \Phi^{i}(x)}{\partial x}\right)
 +\psi_{+}^{j}(x) \Omega_j(\Phi(x))
 -\psi^{l}_{+}\bar{\psi}^{\bar{i}}_{-
}\bar{\psi}^{\bar{j}}_{+}\frac{\partial
 G_{l\bar{j}}}{\partial \bar{\Phi}^{\bar{i}}}\right\}\,.
 \end{equation}
However, this is exactly the supercharge of the $d=2$ generalized
Landau--Ginzburg field theory.

Now, we check how the formula for the central charge
works in this formalism,
\begin{equation}
Z=\sum_{i,x} v^{i,x} \Omega_{i,x}=\int \frac{dW}{d\Phi}
\frac{d\Phi}{dx}dx\,,
\end{equation}
in full agreement with the computation performed
in Sec. 2. One can check
that the zeros of the derivative of the superpotential
of the associated quantum mechanics are nothing but the closed
BPS trajectories!

So, we conclude that the
nonperturbative phenomena that were anticipated in
Sec. 2 actually exist in the form
of the Witten instantonic transition in the associated
${\cal N}=1$ quantum mechanics on the loop space.

\section{Conclusions}

In the theories with the large extra dimensions~\cite{a4} -- a popular 
subject of
theoretical studies at present -- one has to deal with the domain 
walls  on the manifolds of the cylinder type. The issue of the BPS 
saturation  {\it  versus}
nonsaturation of these domain walls is of the paramount importance. 
The dynamical part of this problem obviously reduces to the analysis 
of two-dimensional field theory with the extended supersymmetry 
on  $R\times  S$.

Here we addressed the dynamical question of the existence of the 
BPS  saturated
states within the framework of the generalized Wess--Zumino (or 
Landau--Ginzburg)
models, describing the interaction (possibly, effective) of one or more 
chiral
superfields. Since in such models the central charge $Z\propto \Delta 
W,$
nonvanishing central charges are impossible, at first sight. We 
explained where
this naive point of view is wrong, and presented the theory of the 
BPS 
saturated
states, both at the classical and quantum levels.

We revealed nonperturbative effects lifting the BPS saturation for 
the  classical
BPS solutions. It is shown that at the quantum (nonperturbative) 
level  the BPS
states can exist only if the target space of the Landau-Ginzburg 
model  considered
is noncompact. Using various index-related tools we found the 
number of 
the quantum BPS particles in the theories with one chiral superfield, 
and  the lower bound on this number for more than one chiral 
superfield.

\section{Acknowledgments}

A part of this work was done while one of the authors (M.S.) was 
visiting the Aspen Center for Physics, within the 
framework of the program {\it  Phenomenology
of Superparticles and Superbranes}, in July 1999. Preliminary 
results were reported at this workshop. We would like to 
thank G.~Dvali and G.~Gabadadze for
useful discussions. 

The  work was
supported in part by DOE under the grant number 
DE-FG02-94ER40823.  A.L. acknowledges financial support extended to 
him  during his stay at Theoretical Physics Institute, University of 
Minnesota. The work of A.L. was also supported in part by the
Grant for Support of Scientific Schools number 96-15-96455
and by the RFFI grant number 98-01-00328.

\section*{Appendix: The CFIV Index is Independent of the $D$ 
Terms}
\renewcommand{\theequation}{A.\arabic{equation}}
\setcounter{equation}{0}

The conventional Witten's index is known to be independent
of any small smooth deformations of the supersymmetric theory.
This is not true for the CFIV index.
Still, this index is independent of the continuous variations
of the $D$ terms,
i.e. terms in the Hamiltonian that are
equal to
$$
\{ Q_+, [ \bar{Q}_-, R ]  \} = \{ q_1 ,[q_2 ,R ] \}
$$
or to
$$
\{Q_-, [ \bar{Q}_+, \bar{R} ]  \} =
\{ \bar{q}_2, [ \bar{q}_1 , \bar{R} ] \}
$$
with some operators $R$ and $\bar{R}$. Say, a
small variation of $R$, $\delta H = \{ q_1 ,[q_2 ,\delta R ] \}$, 
results in the
following variation of the  CFIV index:
\begin{eqnarray}
\delta I_{\rm CFIV}&\!\!=&\!\! {\rm tr} \, (-1)^{F} F \exp(-tH)
\{ Q_+, [\bar{Q}_-, \delta R ]  \}(-t)
\nonumber\\[0.2cm]
&\!\!=&\!\! {\rm tr} \, (-1)^{F} F \exp(-tH)
\left(  Q_+ [\bar{Q}_-, \delta R ]  + [\bar{Q}_-, \delta R ] 
Q_+\right)(-t)
\,.
\label{varcf}
\end{eqnarray}
Now, we take $Q_+$  in the second term,
put it to the leftmost position inside the trace, and then drag to the
right, using the fact that
$$
 Q_+ (-1)^F  F  = - (-1)^F  Q_+ F =  - (-1)^FQ_+ - (-1)^F FQ_+\,.
$$
In this way we get
\begin{equation}
\delta I_{\rm CFIV} = - {\rm tr} \,  (-1)^{F} \exp(-tH)
Q_+ [\bar{Q}_-, \delta R ](-t) \,.
\end{equation}
Repeating the same operation with  $\bar{Q}_-$
yields $\delta I_{\rm CFIV}=0$
since $\bar{Q}_+ $ anticommutes with $Q_-$.


\vspace{0.5cm}

\begin{thebibliography}{99}

\bibitem{a1}
V.A.~Rubakov and M.E.~Shaposhnikov,
{\it Phys. Lett.} {\bf 125B}, 136 (1983).

\bibitem{a2}
G.~Dvali and M.~Shifman,
{\it Nucl. Phys.} {\bf B504}, 127 (1997)
[hep-th/9611213].

\bibitem{a3}
G.~Dvali and M.~Shifman,
{\it Phys. Lett.} {\bf B396}, 64 (1997)
[hep-th/9612128];
Erratum {\bf{B407}}, 452 (1997).

\bibitem{a4}
N.~Arkani-Hamed, S.~Dimopoulos and G.~Dvali,
{\it Phys. Lett.} {\bf B429}, 263 (1998)
[hep-ph/9803315];
I.~Antoniadis, N.~Arkani-Hamed, S.~Dimopoulos and G.~Dvali,
{\it Phys. Lett.} {\bf B436}, 257 (1998)
[hep-ph/9804398].

\bibitem{a5}
Z.~Kakushadze, {\it Nucl. Phys.} {\bf B548}, 205 (1999) [hep-
th/9811193];
{\it Nucl. Phys.} {\bf B552}, 3 (1999) [hep-th/9812163]; {\it Nucl. 
Phys.} {\bf
B551}, 549 (1999) [hep-th/9902080]; N. Arkani-Hamed, S.
Dimopoulos, G. Dvali and J. March-Russell, {hep-ph/9811448}; E.
Faraggi and M. Pospelov, {\it Phys. Lett.} {\bf B458}, 237 (1999)
[hep-ph/9901299]; N. Arkani-Hamed and M. Schmaltz, {
hep-ph/9903417};
G. Dvali and A. Smirnov, {hep-ph/9904211}; G. Dvali and G. 
Gabadadze, 
{\it Phys.
Lett.} {\bf B460}, 47 (1999)
[hep-ph/9904221]; L. Hall and C. Kolda, {\it Phys. Lett.} {\bf B459}, 
213 (1999)
[hep-ph/9904236].

\bibitem{a6}
B.~Chibisov and M.~Shifman,
{\it Phys. Rev.} {\bf D56}, 7990 (1997)
[hep-th/9706141];
Erratum {\bf{D58}}, 109901 (1998).

\bibitem{Cecotti:1993rm}
S.~Cecotti and C.~Vafa,
{\it Commun. Math. Phys.} {\bf 158}, 569 (1993)
[hep-th/9211097].

\bibitem{Witten:1978mh}
E.~Witten and D.~Olive,
{\it Phys. Lett.} {\bf B78}, 97 (1978).

\bibitem{Witten:1982dg}
E.~Witten, {\it J. Diff. Geom.} {\bf 17}, 661 (1982).

\bibitem{Cecotti:1992qh}
S.~Cecotti, P.~Fendley, K.~Intriligator and C.~Vafa,
{\it Nucl. Phys.} {\bf B386}, 405 (1992)
[hep-th/9204102].

\bibitem{fi}
P. Fendley and K. Intriligator, {\it Nucl. Phys.} {\bf B372}, 533 
(1992).

\bibitem{Witten:1982df}
E.~Witten,
{\it Nucl. Phys.} {\bf B202}, 253 (1982).

\bibitem{Dvali:1999ht}
G.~Dvali and M.~Shifman,
{\it Phys. Lett.} {\bf B454}, 277 (1999)
[hep-th/9901111].

\bibitem{Cecotti} S.~Cecotti, L.~Girardello, A.~Pasquinucci, {\it Int. J. Mod.
Phys.} {\bf A6}, {2427}
(1991).

\bibitem{Nikita1}
N.~Nekrasov, {\it Four-dimensional holomorphic theories},
PhD Thesis, Princeton University, UMI-9701221.

\bibitem{Nikita2}
N.~Nekrasov, 
{\it Nucl. Phys.} {\bf B531}, 323 (1998) [hep-th/9609219].

\bibitem{BLN}
L.~Baulieu, A.~Losev, and N.~Nekrasov,
{\it Nucl. Phys.} {\bf B522}, 82 (1998)
[hep-th/9707174].

\end{thebibliography}
\end{document}